\documentclass [12pt]{article}
\usepackage{amsmath,amssymb,cite}
\usepackage{appendix}
\usepackage{feynmp}
\usepackage{float}
\usepackage[vcentermath,enableskew]{youngtab}
\usepackage{tabularx}
\usepackage[english]{babel}
\usepackage{graphicx}
\usepackage{indentfirst}
\usepackage{epsfig}
\usepackage{epstopdf}
\usepackage[]{hyperref}
\usepackage[section]{placeins}
\usepackage[stable]{footmisc}
\usepackage{appendix}
\usepackage{tabularx}
\usepackage[english]{babel}
\usepackage{graphicx}
\usepackage{indentfirst}
\usepackage{epsfig}
\usepackage{slashed}
\usepackage{fancyhdr}
\fancypagestyle{plain}{\fancyhead{}
}

\begin{document}

\title{
  Black Hole Interpretation of Quantum Mechanics
}

\author{Badis Ydri\\
  Department of Physics, Annaba University, Annaba, Algeria.
}

\maketitle

\begin{abstract}
Parallels between the measurement problem in quantum mechanics and the black hole information loss problem in quantum gravity are exhibited and then the attempted resolution of the latter in terms of the gauge/gravity duality is extended to the former. 
  
\end{abstract}


Quantum mechanics in the standard Copenhagen interpretation \cite{Bohr}, as formulated originally by von Neumann \cite{VN55},  is characterized by two processes : i) Process I: The collapse of the wave function which occurs during the quantum measurement and ii) Process II: The unitary evolution in time given by the Schrodinger equation.

These two fundamental process can be succinctly illustrated using the Schrodinger's cat experiment \cite{Schrodinger} or its extension the Wigner's friend experiment \cite{Wigner1}. Thus, Schrodinger is taken to play the role of Wigner's friend who performs a quantum measurement on the state of the joint system cat+atom whereas Wigner will perform a quantum measurement on the state of the total joint system Schrodinger+cat+atom.

On the one hand, Schrodinger will provide the Copenhagen description of the system cat+atom which is found to be in a linear coherent superposition of dead-cat/decayed-atom and alive-cat/undecayed-atom until and unless a measurement is performed at which point the state will collapse to one of the said classical alternatives. In other words, Schrodinger provides the local subjective one-person description of reality with direct conscious access to classical pointer states.

On the other hand, the famous/infamous many-worlds description \cite{mw2} is given by Wigner's account who only sees the linear coherent superposition  of sad-observer/dead-cat/decayed-atom and happy-observer/alive-cat/undecayed-atom, i.e. Wigner provides the global objective third-person description of reality who can directly sees the branching worlds when a measurement is performed.

These two descriptions are actually complementarity and not mutually exclusive and this extra principle of complementarity should be thought of as extending the usual complementarity principle between Copenhagen observers due to Bohr \cite{Bohr} to include many-worlds observers as discussed in detail in \cite{Ydri:2020nys}. Of course Nature is seen to be populated only with the first-person local observers of the Copenhagen interpretation but obviously the third-person global observers are not logically prohibited although they are not seen in Nature.

Indeed, the fact that such observers are themselves not directly observed in Nature does not rule them out neither metaphysically nor logically, or even physically \cite{Ydri:2020nys}.

The black hole interpretation of quantum mechanics put forward in a sketchy form in \cite{Ydri:2018ork} relies on  strict physicalism \cite{kim}, i.e. there is no observer-participancy \cite{wheeler} and/or consciousness which is separate from matter \cite{stapp}. This interpretation employs in a crucial way properties of entanglement entropy and falls within the broad scheme of the Copenhagen interpretation (although it does not treat the collapse as a fundamental process but only as an effective description). It is motivated by the ideas of Susskind in \cite{Susskind:2016jjb} who argues among other things for the complementarity relation between the Copenhagen interpretation and the many-world formalism using the properties of entganglement entropy and black holes.

Furthermore, it was pointed out in \cite{Ydri:2018ork} that the complementarity between the local first-person observer of the Copenhagen interpretation (who sees the collapse of the wave function) and the global third-person observer of the many-worlds formalism (who sees directly coherent linear superposition) is the analogue of the complementarity between the asymptotic and infalling observers in black holes.

The inaccessible degrees of freedom, with respect to the local asymptotic observer, in the case of the black hole lie behind the horizon whereas in the case of quantum mechanics the inaccessible degrees of freedom lie behind the Heisenberg cut \cite{Heisenberg} and thus  consist in the degrees of freedom associated with the environment (according to decoherence \cite{zeh}) or with the consciousness of the local first-person observer (within the Wigner-von Neumann interpretation \cite{Wigner1,Ydri:2020nys}). Thus, the collapse of the wave function in the quantum measurement problem  is the analogue of the information loss problem in black holes since they both result from tracing out the inaccessible degrees of freedom yielding a mixed state.

Thus, we have for the quantum measurement problem (where $S$ is the system, $D$ is the detector and $E$ is the environment))

 \begin{eqnarray}
      \rho_e=|\Phi_e\rangle\langle\Phi_e|\longrightarrow \rho_r=\rho_{S+D}=Tr_E|\Psi\rangle\langle\Psi|~,~|\Psi\rangle=U|\Phi_e\rangle|E_0\rangle.
    \end{eqnarray}
Whereas for the information loss problem we have (where ${\cal S}$ is the singularity)

    \begin{eqnarray}
      \rho_{\rm in}=|\psi_{\rm in}\rangle\langle\psi_{\rm in}|\longrightarrow \rho_{\rm out}=\rho_{\rm Hawking}=Tr_{\cal S}|\psi_{\rm out}\rangle\langle\psi_{\rm out}|~,~|\psi_{\rm out}\rangle=S|\psi_{\rm in}\rangle.
    \end{eqnarray} 
However, the information loss problem is only a coarse-graining effect in models of quantum gravity based on the gauge/gravity duality especially the holographic AdS/CFT correspondence \cite{Maldacena:1997re} since  unitarity is guaranteed in these cases in an obvious way.

Similarly, the complementarity between the Copenhagen and many-worlds guarantees that the measurement problem can also be understood as a coarse-graining effect. Together with the fungibility, i.e. the interchangeability  of entanglement, which allows us to compress physical systems, observers and environments into black holes and/or Hawking radiations,  the measurement problem can be recast as an information loss problem \cite{Susskind:2016jjb,Ydri:2018ork}.

 The ER=EPR conjecture \cite{Maldacena:2013xja} in which gravitational Einstein-Rosen bridges or wormholes \cite{Einstein:1935tc} are viewed as Einstein-Podolsky-Rosen-Bell quantum entanglement \cite{EPR35,Bell:1964kc} and vice versa will play an important role.

 For example,  by compressing the cat  and Wigner's friend, i.e. Schrodinger into black holes, it is seen by employing the ER=EPR conjecture, that these black holes will be entangled, and as a consequence there will be quantum Einstein-Rosen bridges between the cat  (first black hole) and Schrodinger (second black hole).

 We can also compress the cat into Hawking radiation and Schrodinger into a black hole or vice versa which allows us to the view the joint system cat+Schrodinger  as an evaporating black hole entangled with its Hawking radiation.

After completed measurement, the Einstein-Rosen bridge is cut at the Schrodinger's end (because he is the one performing the measurement) and thus messages between the cat and Schrodinger can not meet in the wormhole. The snipped ER bridge corresponds to a mixed density matrix and the snipping will cause a firewall between the cat  (viewed as Hawking radiation) and Schrodinger (viewed as an evaporating black hole), i.e. the information loss as characterized by the release of the firewall \cite{Almheiri:2012rt} is exactly seen as the collapse of the state vector as characterized by the snipped ER bridge.

In the many-worlds formalism when Wigner performs his measurement on the system Schrodinger+cat he becomes entangled with it. The system Wigner+Schrodinger+cat is described by an entangled tripartite GHZ state (which is  a generalization of Bell states to three qubits \cite{Bell:1964kc}) in which the cat is entangled with the union of Wigner and his friend \cite{GHZ}.  Thus, when the three systems (Wigner,  Schrodinger and the cat) are compressed into three black holes we get an Einstein-Rosen bridge connecting the three which Susskind also calls the GHZ brane  \cite{Susskind:2016jjb}. This bridge allows messages between the cat and Wigner's friend to be sent to meet inside the GHZ brane in contradiction with the conclusion, using the Copenhagen, which states that messages can not be communicated.

However, because of the separability property of entanglement entropy no messages in the GHZ state can be sent between any two parties since there can be no entanglement between any two of them. So the two interpretations are consistent.

But from this perspective the many-worlds seems more general than the Copenhagen because it captures another very important effect due to the preservation of reversibility and entanglement. There can still be messages sent between Schrodinger and the cat if Schrodinger cooperates with Wigner since the union between any two parties in the GHZ state is entangled with the third party and as a consequence messages can be sent between the union of any two of them and the third one. Hence the two perspectives are consistent and it is in this sense that the many-worlds is complementary to the Copenhagen.

Thus, in principle the measurement problem can be solved (or understood) through the same means used to understand the black hole information loss problem, i.e. by employing  the gauge/gravity duality which maps black holes into gauge theory.

A confirmation, using non-perturbative Monte Carlo methods, of the fact that black holes follow the predictions of the gauge/gravity correspondence is carried out in \cite{Hanada:2013rga} where the gauge theory is given in there case by the celebrated M-(atrix) quantum mechanics, also known as the BFSS  matrix model  \cite{Banks:1996vh}, whereas on the gravity side we have string theory around the black 0-brane (which is a black hole in $10$ dimensions).

Thus, the gauge/gravity duality provides a framework for a novel interpretation of quantum mechanics in which it is seen that the large $N$ limit of  M-(atrix) quantum mechanics becomes given by classical supergravity around a classical black hole given by the 0-brane configuration.  The wave function of the system or more precisely its path integral is given exactly by the classical supergravity action, viz

 \begin{eqnarray}
\Psi=Z_{\rm BFSS}=\exp(-S_{\rm SUGRA}).\label{saddle}
\end{eqnarray}
In other words, there can be no linear coherent superposition of classical states in this limit since the wave function is already fully decohered. Thus, in the large $N$ limit the role of the observer decouples from the unitary evolution and there are no measurement problem and collapse of the wave function.

The supergravity solution is in fact a saddle point of the superstring path integral and in general the BFSS  matrix model  defines the dual gauge theory of the full superstring theory around the black 0-brane. The effect of the observer in the BFSS quantum mechanics side can then be seen as given by a $1/N$ expansion and as such it is  intimately related to the evaporation of the black hole which we will now explain.

First, let us mention for completeness that in the gauge/gravity correspondance \cite{Maldacena:1997re} the two parameters $N$ (rank of the gauge group) and $g_{\rm YM}$ (gauge coupling constant) on the gauge theory (quantum mechanics) side are related to the two parameters $l_s$ (string lenght) and $g_s$ (string coupling constant) on the string theory (black hole and gravity) side as follows 

\begin{enumerate}
\item The gauge theory in the limit $N\longrightarrow\infty$ and $\lambda\longrightarrow\infty$ should be equivalent to classical type IIA supergravity around the 0-brane spacetime. Recall that $\lambda=g_{\rm YM}^2N$ is the 't Hooft coupling constant \cite{tHooft:1973alw} and $\alpha^{\prime}=l_s^2$ is the inverse of the string tension.

\item The gauge theory with $1/N^2$ corrections should correspond to quantum loop corrections, i.e. corrections in $g_s$, in the gravity/string side.

\item The gauge theory with $1/\lambda$ corrections should correspond to stringy corrections, i.e. corrections in $l_s$, corresponding to the fact that degrees of freedom on the gravity/string side are really strings and not point particles.
\end{enumerate}
In the remainder of this letter we will attempt to show that the effect of the observer in the quantum mechanics of   M-(atrix) theory is  intimately related to the evaporation of the black hole of the dual gravity theory of the black 0-brane. 

First, the black hole in this theory is the black 0-brane solution which is  a bound state of $N$ D0-branes (or D-particles) connected by open strings. Indeed, on the gauge theory side the degrees of freedom are given by nine $N\times N$ hermitian matrices $X_I$ whose diagonal elements describe the positions of the $N$ D-particles forming the black hole whereas the off-diagonal elements are fields describing the open strings stretching between these D-particles.


Next, the black 0-brane is a solution of type IIA superstring theory which is when lifted to $11$ dimensions becomes the M-wave solution of M-theory which is a purely geometrical object \cite{Hyakutake:2013vwa2}. Thus, the relevant effective action of M-theory can be computed by concentrating only on the graviton field and demanding local supersymmetry. The effective action of type IIA superstring theory is then obtained via dimensional reduction. Analogously, the near-horizon geometry of the black 0-brane with quantum gravity corrections included is obtained by dimensionally reducing the near-horizon geometry of the M-wave solution.

The thermoynamical properties (temperature, entropy, energy and specific heat) of the black 0-brane can then be computed in the standard way.  For example, the (Hawking) temperature and the specific heat are found to be given by (where $\tilde{U}_0$ indicates the classical horizon in units of  't Hooft coupling $\lambda^{1/3}$)

\begin{eqnarray}
\tilde{T}=a_1\tilde{U}_0^{5/2}(1+\epsilon a_2\tilde{U}_0^{-6}).\label{ht}
\end{eqnarray}
 \begin{eqnarray}
\frac{1}{N^2}\frac{d\tilde{E}}{d\tilde{T}}=\frac{9a_3}{5}\tilde{T}^{9/5}-\frac{3\epsilon a_3a_4}{5}\tilde{T}^{-3/5}.\label{grav0}
\end{eqnarray}
The numerical coefficients $a_i$ can be found in \cite{Hyakutake:2013vwa2} and $\epsilon\sim 1/N^2$. Thus, the specific heat can become negative at low temperatures which means that the black 0-brane behaves as an evaporating Schwarzschild black hole. This instability is obviously removed in the limit $N\longrightarrow\infty$.

This instability of the black hole solution (which corresponds physically to Hawking radiation) is associated with the divergence of the eigenvalues of the matrices $X_I$ as $N$ becomes small. In the BFSS matrix model, this is related to the problem of flat directions or commuting matrices which have zero action in the path integral. A powerful order parameter which exhibits this behavior is given by the so-called extent of space defined by

\begin{eqnarray}
R^2=\frac{1}{N\beta}\int_0^{\beta}dt\sum_{I=1}^9X_I(t)^2.
\end{eqnarray}
The period $\beta$ of the imaginary time $t$ (the time parameter which appears in the BFSS action) is related to the Hawking temperature (\ref{ht}) in the usual way, i.e. $\beta=1/\tilde{T}$.

It is observed in Monte Carlo simulations that the distribution of $R^2$ presents a peak (bound state) and a run-away tail (Hawking instability) \cite{Hanada:2013rga}.  The black hole for small $N$ is therefore only a metastable bound state of the D0-branes and quantum gravity is acting as a destabilizing effect.

This instability can be approximated as follows. By making in the BFSS matrix model the two approximations of 1) quenched determinant (bosonic model) and 2) large number of spatial dimensions $d$ (the Yang-Mills term is approximated by a mass term) we observe at low temperatures (for which the holonomy is exponentially suppressed)  that the BFSS matrix model is a scalar field theory given by the action (see \cite{Ydri:2020fry} and references therein) 

\begin{eqnarray}
S=N\int_{0}^{\infty} dt  {\rm Tr}\big[\frac{1}{2}(\partial_t X_I)^2+\frac{m^2}{2}X_I^2\big]~,~m=d^{1/3}\lambda^{1/3}.\label{har}
\end{eqnarray}
The eigenvalue distribution of any one of the $X_I$ is then given by a Wigner semicircle law with a radius given by
\begin{eqnarray}
R_{\lambda}=\sqrt{\frac{2}{m}}=\sqrt{\frac{2}{d^{1/3}\lambda^{1/3}}}.\label{rad}
\end{eqnarray}
This radius can also be expressed in terms of the expectation value of the extent of space $ R^2$ by the relation

\begin{eqnarray}
R_{\lambda}=\frac{4}{d}\langle R^2\rangle.\label{rad1}
\end{eqnarray}
In summary, as long as we are permitted to use the bosonic approximation (which seems reasonable at low temperatures) and the large dimension expansion (which is quite justified since $d=9$),  it follows that the holonomy of the gauge field is exponentially suppressed at low temperatures and  the dynamics of the BFSS quantum mechanics is given effectively by  (\ref{har}) which describes a matrix harmonic oscillator. As a consequence, we have:

\begin{enumerate}
  \item Strictly speaking the results (\ref{rad}) and (\ref{rad1})  are an $N=\infty$ effect but it is also valid for sufficiently large values of $N$. This clearly vanishes in the limit $\lambda\longrightarrow \infty$ corresponding to the classical black 0-brane solution associated with the saddle point (\ref{saddle}). This represents the fact that in the classical regime the D-particles are absolutely coincident and the black hole does not evaporate Hawking radiation.
  \item  The eigenvalue distribution of  each of the matrix coordinates $X_I$, for large but finite values of $N$ and $\lambda$, is given by a Wigner semicircle law with radius (\ref{rad}) which increases as we decrease $\lambda$. This represents the fact that the D-particles are spread over large distances in the quantum regime thus modeling the Hawking radiation.

    The resulting extent of space $R^2$ given by equation (\ref{rad1}) should then be thought of as an approximation of the distribution of the extent of space  in the full model which presents a peak (bound state) and a run-away tail (Hawking instability) as we decrease $N$  \cite{Hanada:2013rga}.

\item The wave function of the system at low temperatures is the ground state of the matrix harmonic oscillator (or a superposition which is largely dominated by the ground state). It is expected that this picture should hold in the full BFSS model, i.e.  the wave function is expected to be largely dominated by the ground state of the system at low temperatures. At high temperature the partition function is given by the classical result which is identified with the saddle point (\ref{saddle}).
\end{enumerate}
Hence there is no linear coherent superposition in these limits (or they are overwhelmingly dominated by a preferred state) and thus as before the role of the observer decouples from the unitary evolution and there are no measurement problem and collapse of the wave function. The radiated modes away from the black hole are seen to lie at the root of the  superpostion principle and thus they lie at the root of the measurement problem.

\end{document}